\documentclass[preprint,aps,12pt,preprintnumbers,nofootinbib,superscriptaddress] {revtex4}

\usepackage{graphicx}
\usepackage{amssymb}
\usepackage{amsmath}
\usepackage{epstopdf}
\usepackage{subfigure}

\def\ps{p\hskip-0.45em /}

\def\ks{k\hskip-0.45em /}
\DeclareMathOperator{\tr}{tr}

\renewcommand{\r}{\rangle}
\renewcommand{\l}{\langle}

\newcommand{\be}{\begin{eqnarray}}
\newcommand{\ee}{\end{eqnarray}}

\newcommand{\pol}[3]{\varepsilon_{#1 #2 #3}}
\renewcommand{\t}{\tilde}

\renewcommand{\a}{\alpha}
\newcommand{\ad}{{\dot a}}
\renewcommand{\b}{\beta}
\newcommand{\bd}{{\dot b}}

\newcommand{\cd}{{\dot c}}

\newcommand{\dd}{{\dot d}}

\newcommand{\ed}{{\dot e }}

\begin{document}

\title{Amplitudes and Spinor-Helicity in Six Dimensions}

\author{Clifford Cheung}
\affiliation{Department of Physics, Harvard University, Cambridge, MA 02138}
\affiliation{Institute for Advanced Study, School of Natural Sciences, Einstein Drive, Princeton, NJ 08540}

\author{Donal O'Connell}
\affiliation{Institute for Advanced Study, School of Natural Sciences, Einstein Drive, Princeton, NJ 08540}

\date{\today}

\begin{abstract}

The spinor-helicity formalism has become an invaluable tool for
understanding the S-matrix of massless particles in four dimensions. In
this paper we construct a spinor-helicity formalism in six dimensions, and
apply it to derive compact expressions for the three, four and five point
tree amplitudes of Yang-Mills theory. Using the KLT relations, it is a straightforward process to
obtain amplitudes in linearized gravity from these Yang-Mills amplitudes; we demonstrate this 
by writing down the gravitational three and four point amplitudes. Because there
is no conserved helicity in six dimensions, these amplitudes describe the
scattering of all possible polarization states (as well as Kaluza-Klein
excitations) in four dimensions upon dimensional reduction. We also
briefly discuss a convenient formulation of the BCFW recursion relations
in higher dimensions.

\end{abstract}

\maketitle

\section{Introduction}

The spinor-helicity formalism is the natural framework for
representing on-shell scattering amplitudes of massless particles
in four dimensions. This reflects a very basic result from field
theory: asymptotic states of zero mass are uniquely specified by
their momentum and helicity, and as such the S-matrix should be a
function of these variables alone \cite{Wigner:1939cj,Weinberg}.
Unfortunately, this structure is not manifest when amplitudes are
represented using four-vectors and computed with conventional
Feynman diagrams derived from a local action principle. In
particular, for the case of gauge theory and gravity, the cost of
manifest locality and Lorentz invariance is a gauge redundancy
that must be introduced to eliminate extra propagating degrees of
freedom. This gauge freedom implies that the external states are
redundantly labeled by polarization vectors and that the
amplitudes obey non-trivial Ward identities.

In contrast, the spinor-helicity formalism allows us to
write down amplitudes without any mention of gauge symmetry or
polarization vectors. Then, simple considerations of little group
covariance of amplitudes are sufficient to strongly constrain
or even determine the form of on-shell scattering amplitudes
\cite{Benincasa:2007xk,ArkaniHamed:2008gz}. From this point of view, the
framework of spinor-helicity is not merely a computational trick, but is
a way of representing amplitudes in their simplest, most physical form.
Some very nice reviews of the four dimensional spinor-helicity formalism
and its applications can be found in \cite{Dixon:1996wi,Witten:2003nn}.

Until now, there has not been a viable spinor-helicity formalism
in more than four dimensions. There are, nonetheless, many
reasons to suspect that a higher dimensional formalism should be
both elegant and useful. In particular, many of the features of
three and four dimensional spinors reflect their properties
as representations of the SL(2, $\mathbb R$) and SL(2, $\mathbb
C$) Lorentz groups. In six dimensions, the Lorentz group becomes
SL(2, $\mathbb Q$), where $\mathbb Q$ denotes the quaternions
\cite{Kugo:1982bn}, so it seems probable that many of the features
of the familiar four dimensional spinor-helicity variables
have analogues in six dimensions.

In this paper, we construct a spinor helicity formalism in six
dimensions. To orient the reader let us give a flavor of some of
our results. The objects that we will consider are chiral and
anti-chiral six dimensional spinors representing each external
particle. For example, for particle 1, there is an associated
chiral spinor $|1_a \r$, where the $a = 1,2$ index transforms under
one factor of the $\mathrm{SU}(2) \times \mathrm{SU}(2)$ little
group of particle 1. The other $\rm SU(2)$ factor acts on the
$\ad=1,2$ index of an associated anti-chiral spinor $|1_\ad]$.
These little group indices will be ubiquitous in what follows, so
it is worthwhile to comment on them briefly here. While these
indices transform covariantly under the little group, we also know
that they label the basis of physical states in the theory. As such,
any free little group index will ultimately be contracted with
some little group vector that labels the physical polarization of
an external state. This is the point of view that we will adopt
from here on.

Now, the momentum of particle 1 can be expressed as a product of
either chiral spinors or anti-chiral spinors: $-4p_1^\mu = \l 1^a|
\sigma^\mu | 1_a \r = [1^\ad| \tilde \sigma^\mu |1_\ad]$, where
$\sigma$ and $\tilde\sigma$ are the six dimensional Pauli
matrices. These expressions for the momentum in six dimensions
contrast with the four dimensional expression; in that case,
momenta are given by product of one chiral and one anti-chiral
spinor. With the spinors corresponding to particle 2, $|2_b \r$
and $|2_\bd ]$, we can construct a natural Lorentz invariant
object, $\l 1_a | 2_\bd]$, that connects the two particles. The
advantages of this formalism are illustrated by the striking
simplicity of on-shell scattering amplitudes, which we have
computed up to five points. For example, as we will show, the
color-ordered Yang-Mills four point amplitude is given by
\begin{equation}
A_4(1,2,3,4) =  -\frac{i}{st}  \langle 1_a 2_b 3_c 4_d \rangle [
1_\ad 2_\bd 3_\cd 4_\dd ],
\end{equation}
in terms of appropriate quadrilinear contractions of the spinors
associated with each leg. We shall define this contraction in more
detail below. Meanwhile the gravitational four point
function is given by
\begin{equation}
\mathcal M_4(1,2,3,4) = \frac{i}{s t u} \langle 1_a 2_b 3_c 4_d \rangle \langle 1_{a'} 2_{b'} 3_{c'} 4_{d'}  \rangle [  1_\ad 2_\bd 3_\cd 4_\dd ] [ 1_{\ad'} 2_{\bd'} 3_{\cd'} 4_{\dd'} ].
\end{equation}
The amplitude for scattering of a general state, described by some appropriate little group tensor, is found by
contracting the free indices of these expressions against the
little group tensor. Recent work on the $D$-dimensional unitarity
method~\cite{Ellis:2008ir} has some overlap with our results but the
focus of our article is very different.

Upon dimensional reduction to four dimensions, we reproduce the
usual expressions for gauge boson and graviton scattering in four
dimensions. Furthermore, we obtain some four dimensional
amplitudes for scalars: for example, amplitudes that describe
scattering of longitudinal modes of KK vector bosons. From the
gravitational amplitude we can obtain expression for gravitons
scattering with gauge bosons, massive vector bosons and so on.

The structure of the paper is as follows. In section 2 we present
a brief review of the spinor-helicity formalism in four dimensions.
We then go on to develop the six dimensional framework in section 3.
We use this formalism in section 4 to compute beautifully simple
forms for the three point amplitudes in Yang-Mills theory and gravity.
The unique kinematics at three points will require some new ingredients
to express this answer. Section 5 contains some remarks on the BCFW
recursion relations~\cite{Britto:2004ap,Britto:2005fq} and a method for
their efficient use in six dimensions. With this tool we derive the four
point amplitudes in section 6 and the five point Yang-Mills amplitude in section
7, before concluding. The appendices contain some useful identities for
manipulating six dimensional spinors.

\section{A Review of Spinor-Helicity in Four Dimensions}

To begin, let us briefly review the spinor-helicity formalism in
four dimensions. Much of this discussion will have
a direct analogy in six dimensions. The basic point of
spinor-helicity is to represent a light-like four-momentum $p_\mu$
as a bi-spinor
\be p_\mu \sigma^{\mu}_{\alpha\dot{\alpha}} =p_{\alpha
\dot{\alpha}}= \lambda_{\alpha} \tilde{\lambda}_{\dot{\alpha}},
\ee
where $\lambda_\alpha$ and $\tilde{\lambda}_{\dot{\alpha}}$ are
complex valued spinors transforming in the $(2,0)$ and $(0,2)$
representations of the Lorentz group. Since $p_\mu p^\mu = {\rm
det}(p_{\alpha \dot{\alpha}})$ and $p_{\alpha \dot{\alpha}}$ is a
rank one matrix, this bi-spinor represents a null four-vector. In
order to fix $p_\mu$ to be real, we need to impose a reality
condition, $\tilde\lambda = \lambda^*$. However, it is often useful
to analytically continue to complex momenta, so we frequently
relax this condition. The recursion relations
\cite{Britto:2004ap,Britto:2005fq,Badger:2005zh,Cachazo:2005ca,Bedford:2005yy,ArkaniHamed:2008yf,Cheung:2008dn},
which exploit the pole structure in complex momentum space to
recursively relate higher point on-shell amplitudes to lower-point
ones, are a specific instance of this.

While we have specified the momenta in terms of spinors, we know
that a massless particle in four dimensions is labeled not just by
its four-momentum $p_\mu$, but also by its helicity $h=\pm$.
Indeed, in $D$ dimensions it is known that any massless particle
is defined by a ket in a Hilbert space, $|p_\mu, h\r$, where $h$ is
a general label for a linear representation of the
$\mathrm{SO}(D-2)$ little group, the subgroup of the Lorentz group
that leaves $p_\mu$ invariant. Under Lorentz transformations, the
kets transform according to
\be |p_\mu,h\r \rightarrow \sum_{h'} W_{h h'}|
\Lambda_\mu^{\;\;\nu} p_\nu,h'\r \ee
where $\Lambda_\mu^{\;\;\nu}$ and $W_{h h'}$ are Lorentz and
little group transformations, respectively. In four dimensions,
the little group is $\mathrm{SO}(2)$, and so $h$ simply labels the
helicity; then $W_{h h'}$ is a diagonal matrix. For real four
dimensional momenta, $p_{\alpha \dot{\alpha}}$ is manifestly
invariant under
\be \lambda \rightarrow z \lambda , \quad \t \lambda \rightarrow z^{-1} \t \lambda, \ee
where $z$ is a phase for real momenta, or any non-zero complex
number for complex momenta. This is the little group action on
the spinor. With multiple external particles labeled by $i$, each
spinor transforms under its own little group, so $\lambda_i
\rightarrow z_i \lambda_i$. From general considerations
\cite{Benincasa:2007xk} one can show that given helicity
assignments $h_i=\pm$ and spins $s_i=0,1,2$, an on-shell amplitude
transforms as $\mathcal{M} \rightarrow \prod_i z_i^{2s_i h_i}
\mathcal{M}$, which highly constrains the form of amplitudes.

Now that we understand the transformation properties of the
spinors under the Lorentz and little groups, let us comment on Lorentz invariant products. Given two chiral spinors,
$\lambda_{i \alpha}$ and $\lambda_{j \beta}$, there is an obvious
Lorentz invariant product, $\lambda_{i\alpha} \lambda_{j \beta}
\epsilon^{\alpha \beta} \equiv \l \lambda_i \lambda_j \r \equiv \l
ij\r$, and likewise for two anti-chiral spinors,
$\tilde{\lambda}_{i\dot{\alpha}}
\tilde{\lambda}_{j\dot{\beta}}\epsilon^{\dot{\alpha}\dot{\beta}}
\equiv [ \tilde\lambda_i \tilde\lambda_j ] \equiv [ij]$. These
objects are little group covariant since $\l ij \r \rightarrow z_i
z_j \l ij \r$ and $[ij] \rightarrow z_i^{-1} z_j^{-1} [ij]$. All
on-shell amplitudes are functions of these Lorentz invariant,
little group covariant objects. For example, the three point
function of a theory of spin $s$ particles is
\begin{align}
\label{eq:4dYM}
A_3(1^-,2^-,3^+) = \left(\frac{\l 1 2 \r^3}{\l 23 \r \l 31 \r}
\right)^s,&&\quad A_3(1^+,2^+,3^-) = \left(\frac{[ 1 2 ]^3}{[ 23 ]
[
31 ]} \right)^s , \\
\label{eq:4dF3}
A_3(1^-,2^-,3^-) = \frac{1}{M^2} \left(\l 12 \r \l 23 \r \l 31 \r
\right)^s,&&\quad A_3(1^+,2^+,3^+) = \frac{1}{M^2}\left([12][23][31]
\right)^s
\end{align}
with no reference to polarization vectors. In Eq.~\eqref{eq:4dF3} we
have included factors of $1/M^2$ on dimensional grounds; these amplitudes
vanish in the case of pure Yang-Mills theories but arise from a dimension
six operator $\tr F_{\mu \nu} F_\rho{}^\nu F^{\mu \rho}$ in an effective
theory. In fact, Eq.~\eqref{eq:4dYM} and Eq.~\eqref{eq:4dF3} are the
form of the three point amplitude to all orders in perturbation theory,
as a consequence of momentum conservation and little group covariance
\cite{Benincasa:2007xk,ArkaniHamed:2008gz}. Beginning with this
three point amplitude, the BCFW recursion relations can then be used to
construct all higher point functions from these amplitudes.

That said, if we wish to make a direct connection to more
conventional methods for computing amplitudes, then we can still
define polarization vectors in terms of four dimensional spinors.
Consider a particle of momentum $p$; it is convenient to denote
the associated spinors as $\lambda = |p \rangle$ and $\tilde \lambda
= |p]$. Then the polarization vectors associated with this particle
can be written as
\be \varepsilon_{-}^\mu &=& \frac{1}{\sqrt 2} \frac{\l p|
\sigma^\mu |q]}{ [pq]}\\
\varepsilon_+^\mu &=& \frac{1}{\sqrt 2} \frac{\l q |\sigma^\mu
|p]}{ \l pq\r}, \ee
where $| q \r$ and $|q]$ are reference spinors. Note that the
polarization vectors are appropriately covariant under the little
group of $|p \rangle$ and $|p]$, but are manifestly
invariant under little group transformations acting on the
reference spinors.

\section{Constructing Spinor-Helicity in Six Dimensions}

It is straightforward to extend the construction of the previous
section to six dimensions. Our goal is to construct a spinor
representation of the momentum, $p_\mu$, that transforms
appropriately under the Lorentz and little groups. In particular,
since the Lorentz group is $\mathrm{SO}(6)\simeq \mathrm{SU}(4)$,
these six dimensional spinors are complex four component objects,
transforming in the fundamental of $\mathrm{SU}(4)$ under Lorentz
transformations. Since the antisymmetric representation of
$\mathrm{SU}(4)$ is six dimensional, we expect $p_\mu$ to be
written as some antisymmetric product of two spinors. Moreover,
since the little group is $\mathrm{SO}(4)\simeq
\mathrm{SU}(2)\times \mathrm{SU}(2)$ for real momenta,
then the spinors should have two $\mathrm{SU(2)}$ spinor
indices. For the purposes of this paper we consider complex
momenta, for which the associated spinors need not satisfy any
reality conditions. Consequently, the little group is extended to
$\mathrm{SL}(2,\mathbb{C})\times \mathrm{SL}(2,\mathbb{C})$.

\subsection{From the Dirac Equation}

Indeed, solutions of the Dirac equation for a null momentum, $p_\mu$,
have the properties we require, as we will now see. The equations we must solve are
\begin{equation}\label{eq:dirac}
p_\mu  \sigma^\mu_{\;\;AB} \lambda^B = 0, \;\;\; p_\mu \tilde
\sigma^{\mu AB} \tilde \lambda_B = 0,
\end{equation}
where $\sigma^\mu_{\;\;AB}$ and $\tilde\sigma^{\mu AB}$ are
six dimensional antisymmetric Pauli matrices
described in detail in Appendix~\ref{sec:clifford}.
Our choice of basis is such that the $\sigma$ matrices restricted to $\mu = 0,1,2,3$ reduce to a familiar (Weyl) choice of $\gamma$ matrices in four dimensions:
\begin{align}
\label{eq:4dsigma} \sigma^\mu &= \begin{pmatrix}
0 & {}^{(4)}\sigma^{\mu \alpha}{}_{\dot \alpha} \\
-{}^{(4)}\sigma^{\mu \alpha} {}_{\dot \alpha}{}^T & 0
\end{pmatrix},
\;\;\;\; \tilde \sigma^\mu = \begin{pmatrix}
0 & {}^{(4)}\sigma_{\mu \alpha}{}^{\dot \alpha} \\
-{}^{(4)}\sigma_{\mu \alpha} {}^{\dot \alpha}{}^T & 0
\end{pmatrix},
\;\;\;\; \mu = 0,1,2,3,
\end{align}
where ${}^{(4)}\sigma^\mu{}_{\alpha \dot \alpha}  = (\sigma_0,
\sigma_1, \sigma_2, \sigma_3)$ are the usual four dimensional sigma
matrices, and $\alpha, \; \dot \alpha = 1,2$ are spinor indices of
the four dimensional Lorentz group.
We take $\lambda^A$ and $\tilde\lambda_A$, the solutions of
Eq.~\eqref{eq:dirac}, to be in the fundamental and
anti-fundamental representations of $\mathrm{SU}(4)$, respectively. Unlike the
familiar case of $\mathrm{SU}(2)$, the fundamental and
anti-fundamental representations of $\mathrm{SU}(4)$ are
inequivalent since there is no tensor that can raise or lower
indices. In fact, the only non-trivial invariant tensor is a four
index object, $\epsilon_{ABCD}$.

Since $p_\mu \sigma^\mu_{\;\;AB}$ is a rank two matrix, there is a
two dimensional space of solutions for the $\lambda$ equation in
Equations~\eqref{eq:dirac} that we can label by $a = 1, 2$. We do the
same for the $\tilde\lambda$ equation, labeling by $\dot{a}=1,2$.
Thus, the chiral and anti-chiral spinors can be written as
$\lambda^{Aa}$ and $\tilde \lambda_{A \ad}$~\footnote{Including the little group label, each chiral and anti-chiral spinor is a four by two matrix.
Consequently, we can reinterpret these objects as quaternionic two-component
spinors, where each quaternion is represented by a two by two
matrix. This reflects a fact we alluded to earlier: the Lorentz group in six
dimensions is isomorphic to $\mathrm{SL}(2,\mathbb{Q})$~\cite{Kugo:1982bn}.}~.
We will
see that these $a$ and $\dot{a}$ indices are precisely the
$\mathrm{SU}(2)\times \mathrm{SU}(2)$ indices of the little group.

If the momentum $p$ happens to lie in the privileged four-space
fixed by our choice of the $\sigma$ matrices, that is $p = (p^0,
p^1, p^2, p^3, 0,0)$, then we can choose solutions of
Eq.~\eqref{eq:dirac} given by
\begin{equation}
\lambda^A_{\;\;a} = \begin{pmatrix}
0 & {}^{(4)}\lambda_\alpha \\
 {}^{(4)}\tilde \lambda^{\dot \alpha} & 0
\end{pmatrix}, \;\;\;
\tilde \lambda_{A \ad} = \begin{pmatrix}
0 &  {}^{(4)}\lambda^\alpha \\
 {}^{(4)}\tilde \lambda_{\dot \alpha} & 0
\end{pmatrix} ,
\end{equation}
where $ {}^{(4)}\lambda$ and $ {}^{(4)}\tilde \lambda$ are four
dimensional spinors. Note the position of the four dimensional
spinor indices; these follow from the positions of the indices in
Eq.~\eqref{eq:4dsigma}.

In computations, it is frequently convenient to use a bra-ket
notation, and so we write
\begin{equation}
\lambda^{a} = |p^a\r, \quad \t \lambda_\ad = | p_\ad].
\end{equation}
When several particles scatter we will choose to label the kets by
the label of the particle for brevity. It is possible to normalize
the basis of spinor solutions so that
\be
p_\mu \tilde \sigma^{\mu AB} &=& p^{AB} = \lambda^{Aa}
\lambda^{Bb}\epsilon_{ab}  = |p^a\r \epsilon_{ab} \l p^b|, \\
p_\mu \sigma^\mu_{\;\;AB} &=& p_{AB} = \tilde \lambda_{A\dot{a}} \tilde \lambda_{B \dot b}
\epsilon^{\dot{a} \dot{b}}
= |p_\ad ]\epsilon^{\dot{a} \dot{b}} [p_\bd|  .
\ee
With the help of Eq.~\eqref{eq:sigmaprops}, we can express the
momentum vector itself in terms of the spinors as \be p_\mu &=&
-\frac{1}{4} \l p^{ a}| \sigma_{\mu}| p^{b}\r \epsilon_{ab} =
-\frac{1}{4} [ p_{\dot{a}} |  \tilde \sigma_\mu| p_{\bd}]
\epsilon^{\ad \bd}. \ee From this point on, we will freely raise
and lower the $\mathrm{SU}(2)$ indices, $a$ and $\dot a$, using
the definitions
\begin{subequations}
\begin{align}
|p_a\r &= \epsilon_{a b} |p^b\r, \\
|p^\ad] &= \epsilon^{\ad \bd} |p_\bd].
\end{align}
\end{subequations}
We define $\epsilon^{12}=1$ and $\epsilon_{12}=-1$.

\subsection{To the Little Group}

Earlier, we remarked that the $a$ index of the spinor $|p_a\r$
transforms under the little group. Let us now take a moment to
explain why this is so. Consider a Lorentz transformation
$\Lambda$ with the property that $\Lambda^\mu{}_\nu p^\nu =
p^\mu$; that is, $p$ is invariant under the transformation. Then
$\Lambda$ is an element of the SO(4) little group associated with
$p$. This transformation acts on the spinor $\lambda$ by a unitary
matrix $U$ and upon $\tilde \lambda$ by the inverse matrix
$U^{-1}$. If we define $\lambda' = U \lambda$, then $\lambda'$
satisfies the Dirac equation since
\begin{equation}
p_\mu  \sigma^\mu \lambda' =
 p_\mu  (U^{-1} U \sigma^\mu U \lambda) =  \Lambda^{\;\;\mu}_{\nu} p_\mu
( U^{-1} \sigma^\nu \lambda) = U^{-1} (p_\mu  \sigma^\mu
\lambda)=0.
\end{equation}
Consequently, we may write $\lambda'_a = M_a{}^b \lambda_b$ for
some matrix $M$, as the spinors $\lambda_a$ form a basis for the
solution space. Using the two expressions for $\lambda'$ it is straightforward
to show that
\begin{equation}
-\frac{1}{4} \lambda^{\prime a} \sigma^\mu \lambda'_a
= p^\mu
=p^\mu \det M
\end{equation}
Therefore we conclude that $M \in \mathrm{SL}(2, \mathbb C)$.
Similarly, the spinors $\tilde \lambda$ transform as $\tilde
\lambda' = \tilde M \tilde \lambda$ where $\tilde M \in
\mathrm{SL}(2, \mathbb C)$. Since there is in general no relation
between $M$ and $\tilde M$ we conclude that the full space of
transformations that leave the momentum invariant, i$.$e$.$ the
little group, is $\mathrm{SL}(2, \mathbb{C}) \times \mathrm{SL}(2,
\mathbb{C})$.

\subsection{Invariants and Covariants}

In analogy with the four dimensional case we can now construct a
set of natural Lorentz invariant, little group covariant objects~\footnote{
The language here may seem a bit odd, because the little
group is by definition the subgroup of the Lorentz group that
leaves $p_\mu$ invariant. Thus we should expect that anything
Lorentz invariant is little group invariant as well. While this
is certainly true, the little group can also be understood as a
separate set of transformations that acts on and defines the basis
of polarizations for {\it each} external particle. For example,
for a massive particle in four dimensions, the little group is
SO(3)---thus, while an SO(3) little group index is of course
rotated via boosts, it can also be thought of as an index that is
to be contracted with some three-vector polarization built out of
the basis polarizations. Throughout this paper, our view is that
these indices label these physical polarizations of external
states. Thus, when we say some object is Lorentz invariant but
little group covariant, we mean a genuine Lorentz invariant, which
happens to depend on the polarization states of the various particles
scattering.}.
Lorentz invariant contractions of spinors associated to two particles
labeled by $i$ and $j$ are
\begin{equation}
\langle i^a | j_\bd] = \lambda_{i}^{\;Aa} \tilde \lambda_{j A \bd} = [j_\bd | i^a \rangle,
\end{equation}
which is a two by two matrix that transforms in the bifundamental
under a separate $\mathrm{SU}(2)$ little group factor for particle
$i$ and for particle $j$. For spinors associated with momenta $p$
and $q$ in the privileged four-space of our $\sigma$ matrices, we
find
\begin{align}
\langle i_a | j_\bd]
&=
 \begin{pmatrix}
-[ij] & 0 \\
0 & \l ij \r
\end{pmatrix}_{a \bd}, \\
[ i_\ad | j_b \r
&=
 \begin{pmatrix}
[ij] & 0 \\
0 & - \l ij \r
\end{pmatrix}_{\ad b}.
\end{align}
Note that $\det [i | j \r = -2 p_i \cdot p_j$.
In addition, using the four index antisymmetric tensor, we can
construct a Lorentz invariant from four spinors labeled by
$i,j,k,l$:
\be \l i^a j^b k^c l^d \r &=& \epsilon_{ABCD}\lambda_i^{\;Aa}
\lambda_j^{\;Bb}\lambda_k^{\;Cc}\lambda_l^{\;Dd} \\
 \left[ i_{\dot{a}} j_{\dot{b}} k_{\dot{c}} l_{\dot{d}}\right] &=&
\epsilon^{ABCD}\tilde\lambda_{i A \ad}\tilde \lambda_{j B
\bd}\tilde \lambda_{k C \dot{c}}\tilde \lambda_{l D \dot{d}}. \ee
Finally, given particles labeled by $i$, $j$, and $k_1,\ldots,
k_{2n+1}$, we define
\begin{align}
\langle i_a | \ps_{k_1} \ps_{k_2} \cdots \ps_{k_{2n + 1}} | j_b
\rangle &= \lambda_{i\;\; a}^{\;A_1} (p_{k_1} \cdot \sigma_{A_1
A_2}) (p_{k_2} \cdot \tilde \sigma^{A_2 A_3})
\cdots (p_{k_{2n+1}} \cdot \sigma_{A_{2n+1} A_{2n+2}}) \lambda_{m\;\; b}^{\;\;A_{2n+2}} \\
\langle i_a | \ps_{k_1} \ps_{k_2} \cdots \ps_{k_{2n}} | j_\bd ] &=
\lambda_{i\;\; a}^{\;A_1} (p_{k_1} \cdot \sigma_{A_1 A_2})(
p_{k_2} \cdot \tilde \sigma^{A_2 A_3} )\cdots (p_{k_{2n+1}} \cdot
\tilde \sigma^{A_{2n} A_{2n+1}}) \tilde \lambda_{j  A_{2n+1} \bd}
.
\end{align}

\subsection{Polarization Vectors}

The advantage of the spinor-helicity formalism is that the
polarization states of the external particles live in irreducible
representations of the little group. In contrast, conventional
Feynman diagrammatics forces us to represent polarization states
redundantly as Lorentz six-vectors. In this section we make contact
with this picture by writing polarization six-vectors in
terms of the six dimensional spinors.

To begin, we pick a
reference six-vector $q$ such that $p \cdot q \neq 0$, where $p$
is the particle momentum. Associated with $q$ are two spinors
such that $q = |q^{a}\r \l q^{b} |\epsilon_{ab}$ and $q =
|q_{\ad}][q_{\bd}| \epsilon^{\ad \bd}$. We then define the
polarization vectors to be
\be \varepsilon^\mu_{\;\;a \ad} &=&
\frac{1}{\sqrt 2} \; \l p_a |\sigma^\mu| q_b\r \;
(\l q_b |p^\ad])^{-1} \\
&=& \frac{1}{\sqrt 2} \; (\l p^a |q_\bd])^{-1} \;
[q_\bd| \tilde \sigma^\mu| p_\ad].
\ee
We note that, in contrast to the four dimensional case, the
polarizations are not simply labeled by helicity + or -, but by
$\mathrm{SO}(4)\simeq \mathrm{SU(2)}\times \mathrm{SU(2)}$ little
group indices.
On the other hand, just as in four dimensions, a little group transformation
acting on the reference spinors $|q \r$ and $|q]$ has no
effect on the polarization. We have normalized the polarization
vectors so that
\begin{equation}
\varepsilon^\mu_{\;\;a \ad} \varepsilon_{\mu b \bd} = \epsilon_{a
b} \epsilon_{\ad \bd}.
\end{equation}

On physical grounds, the polarization vectors must satisfy two key
properties: they must transform appropriately under gauge
transformations, and furthermore they must form a complete set of
vectors transverse to the momentum $p$. Let us demonstrate that
our vectors satisfy these requirements, starting with the former.
Choose a new gauge $q'$ such that $p \cdot q' \neq 0$; associated
with this new gauge are new spinors $|q'\r$. In general, we can
write
\begin{equation}
|q'_c \r = A_c{}^b |q_b \r + B_c{}^a |p_a \r,
\end{equation}
where $[p_\ad| q'_c\r = A_c{}^b \l q_b |p_\ad ]$.
Now, since $\det \, [ p | q \r = -2 p \cdot q \neq 0$ and similarly
$\det \,[p | q' \r \neq 0$, it follows that $\det A \neq 0$ so
that $A$ is an invertible matrix. Using the definition of the polarization
vectors, it is now a straightforward calculation to show that
\begin{equation}
\varepsilon_{\;\;\;a \ad}^{\prime \mu} = \varepsilon^\mu_{\;\;a \ad}
+ \Omega_{a \ad} p^\mu ,
\end{equation}
where
\begin{equation}
\Omega_{a \ad} = - \sqrt 2 (A_c{}^b \l q_b| p^\ad] \,)^{-1} B_{c a}.
\end{equation}
Thus, the polarization vectors shift under a gauge transformation by an amount
proportional to the associated momentum, as desired. Finally, it is a straightforward
computation to show that the polarization vectors form a complete set in the sense
that
\begin{equation}
\varepsilon_{\;\;a \ad}^\mu \varepsilon^{\nu a \ad} = \eta^{\mu
\nu} - \frac{1}{p \cdot q} (p^\mu q^\nu + p^\nu q^\mu).
\end{equation}

\section{The Three Point Function}

In this section, we derive compact forms for the three point
scattering amplitudes of Yang-Mills theory and gravity. It will
actually be illuminating to first try and guess the form of the
three point amplitude directly from little group considerations
alone. In particular, given particles 1, 2, and 3, with little group
indices $(a,\ad)$, $(b,\bd)$, and $(c,\cd)$, we know that the
amplitude must have exactly one of each index. The most obvious
guess is $ \l 1_a |2_\bd] \l 2_b | 3_\cd] \l 3_c |1_\ad] /M^2$,
where some scale $M$ has been included on dimensional grounds.
As it turns out, this amplitude arises
precisely from the higher dimension operator $\tr F_{\mu \nu}
F^{\nu\rho} F_{\rho}^{\;\;\mu} / M^2$ which can be added to
Yang-Mills theory. The four dimensional analogs of this amplitude
are $A_3(1^- 2^- 3^-)$ and $A_3(1^+ 2^+ 3^+)$. If we are
concerned with the renormalizable couplings of Yang-Mills theory,
then no such scale $M$ is present, and moreover momentum
conservation forces all the kinematic invariants $p_i \cdot p_j$
to vanish. Thus, dimensional analysis tells us that to write down
the three point amplitude for Yang-Mills theory, it will be
necessary to invert the quantities $ \l i| j]$. However, this is
naively a problem because $\det \l i | j] = -2 p_i \cdot p_j = 0$.
In this way, we see that a new ingredient is necessary.

The solution to this problem is as follows.
Since $\l i_a | j_\bd]$ is a rank one matrix, it can be expressed
as a product of two two-component objects, $u_{ia}$ and $\tilde
u_{j\bd}$, such that $\l i_a |j_\bd ] = u_{ia} \t u_{j\bd}$. Since
$u_i$ and $\t u_j$ are quite reminiscent of four dimensional
spinors, we know the natural inversion of these $u$'s. In
particular, we introduce spinors $w_i$ and $\tilde w_j$ such that
$u_i^a w_{ia} = 1$ and $\t u_j^a \t w_{ja} = 1$. Of course, these
inverses are not uniquely defined, and we will discuss this
ambiguity in detail below. However, we ultimately find a
factorized form for the Yang-Mills three point function given
by
\begin{equation}
A_3(1_{a\ad},2_{b\bd},3_{c\cd}) = i \Gamma_{abc} \tilde \Gamma_{\ad
\bd \cd},
\end{equation}
where the tensors $\Gamma$ and $\tilde \Gamma$ are simply
\begin{align}
\Gamma_{abc} &= u_{1a} u_{2b} w_{3c} +u_{1a} w_{2b} u_{3c} +w_{1a} u_{2b} u_{3c} ,\\
\tilde \Gamma_{\ad \bd \cd} &= \t u_{1\ad} \t u_{2\bd} \t w_{3 \cd} + \t u_{1\ad} \t w_{2\bd} \t u_{3 \cd} + \t w_{1\ad} \t u_{2\bd} \t u_{3 \cd} .
\end{align}
We will frequently use notation like $A_n(1_{a\ad} , 2_{b\bd}, \ldots)$ to
indicate an $n$ point gauge theory amplitude where the little group indices
of particle one are $(a, \ad)$ and so on.

\subsection{Three Point Amplitude in Yang-Mills}

Let us now discuss these issues in detail. It is helpful for the purposes of clarity to choose
$p_1$, $p_2$ and $p_3$ in the privileged four-space of our choice
of the $\sigma$ matrices. Then the Lorentz invariant brackets can
be taken to be of the form
\begin{equation}
\langle 1_a | 2_\bd ] = \begin{pmatrix}
-[12] & 0 \\
0 & \langle 12 \rangle
\end{pmatrix}.
\end{equation}
From our experience with the three point function in four
dimensions, we know that either $\langle 12 \rangle = 0$ or
$[12]=0$. We suppose that $[12]=0$. Thus, the Lorentz invariants
are of the form
\begin{equation}
\langle 1_a | 2_\bd ] = \begin{pmatrix}
0 & 0 \\
0 & \langle 12 \rangle
\end{pmatrix}.
\end{equation}
Now we can define two component vectors $u_i$ and $\tilde u_i$ for
$i=1,2,3$ such that the equations
\begin{subequations}
\label{eq:defus}
\begin{align}
\langle 1_a | 2_\bd] &= u_{1a} \tilde u_{2 \bd} & \langle 1_a | 3_\cd] &= -u_{1a} \tilde u_{3 \cd} \\
\langle 2_b | 3_\cd] &= u_{2b} \tilde u_{3 \cd} & \langle 2_b | 1_\ad] &= -u_{2b} \tilde u_{1 \ad} \\
\langle 3_c | 1_\ad] &= u_{3c} \tilde u_{1 \ad} & \langle 3_c | 2_\bd] &= -u_{3c} \tilde u_{2 \bd}
\end{align}
\end{subequations}
hold. In terms of our choice of spinors, we can choose $u_i = (0,
N_i)$ and $\tilde u_i = (0, \tilde N_i)$; then the solution of the
Equations~\eqref{eq:defus} can be written as
\begin{equation}
N_2 = \frac{\langle 23 \rangle}{\langle 31 \rangle} N_1, \;\;\;
N_3 = \frac{\langle 23 \rangle}{\langle 12 \rangle} N_1, \;\;\;
\tilde N_1 = \frac{\langle 12 \rangle \langle 31 \rangle}{\langle 23 \rangle} \frac{1}{N_1}, \;\;\;
\tilde N_2 = \frac{\langle 12 \rangle}{N_1}, \;\;\;
\tilde N_3 = \frac{\langle 31 \rangle}{N_1}.
\end{equation}
More general solutions of Equations~\eqref{eq:defus} can be obtained by
little group transforming this explicit solution. Notice that the
overall normalization of these SU(2) spinors $u$ and $\t u$ is not
determined, but that a change in normalization of the $u_i
\rightarrow N u_i$ has the opposite effect on the $\tilde u_i
\rightarrow 1/N \tilde u_i$.

We can establish a key property of the $u$ and $\tilde u$ spinors
by studying conservation of momentum. In spinorial terms, momentum
conservation reads
\begin{equation}
| 1^\ad ] [ 1_\ad | + |2^\bd ][ 2_\bd | + | 3^\cd ] [ 3_\cd| = 0 = | 1^a \rangle \langle 1_a | + | 2^b \rangle \langle 2_b | + | 3^c \rangle \langle 3_c |.
\end{equation}
Consider contracting the first half of this statement with $\langle 1_a|$. We find
\begin{equation}
0= \langle 1_a | 2^\bd] [ 2_\bd| + \langle 1_a | 3^\cd] [ 3_\cd|
= u_{1 a} \tilde u_2^\bd [2_\bd| - u_{1a} \tilde u_3^\cd [3_\cd|,
\end{equation}
so that $\tilde u_2^\bd [2_\bd| = u_3^\cd [3_\cd|$. Similarly we find $\tilde u_1^\ad [1_\ad| = u_3^\cd [3_\cd|$ and that $u^a \langle i_a | = u^a \langle j_a |$ for all $i, j$.
Since we will frequently encounter little group contractions such as $u^a \l 1_a|$ in the following, we will denote them as $\l u \cdot 1|$.

As we mentioned earlier in this section, the next ingredient we need is an
inverse of each of the SU(2) spinors $u_i$ and $\tilde u_i$. We
define $w_i$ and $\tilde w_i$ so that
\begin{equation}
u_{i a} w_{i b} - u_{i b} w_{i a} = \epsilon_{a b}, \;\;\;
\tilde u_{i \ad} \tilde w_{i \bd} - \tilde u_{i \bd} \tilde w_{i \ad} = \epsilon_{\ad \bd}.
\end{equation}
for all $i$. The $w_i, \tilde w_i$ are not uniquely specified; given one
choice of $w_i$, for example, then the choice $w'_i = w_i + b_i u_i$ is
equally good. We will reduce this $b$ redundancy a little, but we will
not fully eliminate it. The additional constraint we impose is motivated
by conservation of momentum, which can now be written in the form
\begin{equation}
|1 \cdot u \rangle \left( \langle w_1 \cdot 1| + \langle w_2 \cdot 2| + \langle w_3 \cdot 3| \right)
- \left(|w_1 \cdot 1 \rangle + |w_2 \cdot 2 \rangle + |w_3 \cdot 3 \rangle \right) \langle u_1 \cdot 1| =0.
\end{equation}
We impose the stronger equation
\begin{equation}
\label{eq:split}
|w_1 \cdot 1 \rangle + |w_2 \cdot 2 \rangle + |w_3 \cdot 3 \rangle = 0.
\end{equation}
There is still residual redundancy: we can shift $w_i \rightarrow w_i +
b_i u_i$ where $b_1+b_2+b_3 =0$. In view of this remaining redundancy,
it is interesting to ask what tensors we can construct from the $u$'s
and the $w$'s which are invariant under a $b$ change. It is easy to see
that the quantities
\begin{align}
\Gamma_{a b c} &= u_{1a} u_{2b} w_{3c} + u_{1a} w_{2b} u_{3c} + w_{1a} u_{2b} u_{3c} , \\
\tilde \Gamma_{\ad \bd \cd} &= \tilde u_{1\ad} \tilde u_{2\bd} \tilde w_{3 \cd} + \tilde u_{1 \ad} \tilde w_{2 \bd} \tilde u_{3 \cd } +\tilde w_{1 \ad} \tilde u_{2 \bd} \tilde u_{3 \cd}
\end{align}
are invariant; for example, under a $b$ shift, $\Gamma$ shifts by $\sum_i
b_i u_{1a} u_{2b} u_{3c} = 0$. In addition, the quantity $\Gamma \tilde
\Gamma$ is unambiguously normalized. To gain some more intuition for
the physical meaning of these objects, let us examine them in terms of
the explicit solution we have obtained for the $u$'s. From the definition
of $w_i$ we find that
\begin{equation}
w_{ia} =
\begin{pmatrix}
\frac{1}{N_i} \\
b_i N_i
\end{pmatrix}, \quad (\mbox{no sum on } i)
\end{equation}
so that Eq.~\eqref{eq:split} becomes
\begin{equation}
\label{eq:splitExplicit}
\begin{pmatrix}
- \frac{1}{N_1} \ell_1 - \frac{1}{N_2} \ell_2 -  \frac{1}{N_3} \ell_3 \\
b_1 N_1 \t \ell_1 + b_2 N_2 \t \ell_2 + b_3 N_3 \t \ell_3
\end{pmatrix}
=
\begin{pmatrix} 0 \\ 0 \end{pmatrix}.
\end{equation}
Upon substitution of the explicit solutions for the $N_i$, and use of
the equations
\begin{equation}
\t \ell_2 = \frac{\langle 31 \rangle}{\langle 23 \rangle} \t \ell_1, \quad \t \ell_3 = \frac{\langle 12 \rangle}{\langle 23 \rangle} \t \ell_1,
\quad
\ell_1 + \frac{\langle 31 \rangle}{\langle 23 \rangle} \ell_2 +  \frac{\langle 12 \rangle}{\langle 23 \rangle} \ell_3 = 0,
\end{equation}
which follow from conservation of momentum in the four dimensional
formalism, we see that Eq.~\eqref{eq:splitExplicit} is satisfied when $b_1
+ b_2 + b_3 = 0$, as anticipated. As for the quantity $\Gamma \t \Gamma$,
let us content ourselves with an examination of one component. For
example, we find that
\begin{equation}
\Gamma_{221} \t \Gamma_{221} = \frac{N_1 N_2}{N_3}  \frac{\t N_1 \t N_2}{\t N_3} = \frac{\langle 12 \rangle^3}{\langle 23 \rangle \langle 31 \rangle}.
\end{equation}
Notice that this is proportional to $A_3(1^- 2^- 3^+)$ from four dimensional
Yang-Mills theory. Our next task will be to see why this is so.

We begin in familiar territory. The usual color-ordered amplitude in non-Abelian gauge theory is given in terms of polarization vectors by
\begin{equation}
A_3 = \frac{i}{\sqrt 2} \left( [\varepsilon_{1 a \ad} \cdot
\varepsilon_{2 b \bd} ][\varepsilon_{3 c \cd} \cdot (p_1 - p_2)] +
[\varepsilon_{2 b \bd} \cdot \varepsilon_{3 c \cd}][
\varepsilon_{1 a \ad} \cdot (p_2 - p_3)] + [\varepsilon_{3 c \cd}
\cdot \varepsilon_{1 a \ad}][ \varepsilon_{2 b \bd} \cdot (p_3 -
p_1)] \right).
\end{equation}
We must rewrite this amplitude in a manifestly gauge-invariant form. A key observation is that
\begin{equation}
\pol 1 a \ad \cdot p_2 = -\frac{1}{\sqrt 2} u_{1 a} \t u_{1 \ad}.
\end{equation}
Inner products of two polarization vectors are not gauge invariant so we do not expect a simple expression for such inner products. We will choose the same gauge $\mu$, $\t \mu$ for all three particles. Then, from the spinorial definitions of the polarization vectors, we find that
\begin{align}
\pol 1 a \ad \cdot \pol 2 b \bd &= - \langle 1_a | \t \mu_{\ed} ][\t \mu_{\ed}| 2^\b \rangle^{-1} [2_\bd | \mu_d\rangle \langle \mu_d | 1^\ad]^{-1} \\
&= - \langle 2_b | \t \mu_{\ed} ][\t \mu_{\ed}| 1^\a \rangle^{-1} [1_\ad | \mu_d\rangle \langle \mu_d | 2^\bd]^{-1},
\end{align}
with similar equations holding for the other inner products. 
From this point, one systematically uses the definitions of the $u$'s and the $w$'s to remove all of the matrices from the expression for the amplitude. After some work, we find the desired result: the amplitude is
\begin{equation}
A_3(1_{a \ad},2_{b \bd},3_{c \cd})= i \Gamma_{a b c} \t
\Gamma_{\ad \bd \cd}.
\end{equation}
We could now continue to express the amplitude in terms of spinor
contractions $\langle i | j]$ and their appropriately defined
pseudo-inverses; however, we find it to be more convenient not to
do this.

\subsection{Three Point Amplitude in Gravity}

Next, let us consider the three point function in linearized
gravity. From the point of view of the SO(4) little group, the graviton
polarization tensor is a traceless, symmetric tensor. That is, writing
the polarization tensor as $\varepsilon^\mu_{mn}$ where $\mu = 0, \ldots 5$ is
a six dimensional Lorentz index and $m,n = 1, \ldots 4$ are vector
indices of SO(4), we know that the equations $\varepsilon_{mn} = \varepsilon_{nm}$
and $\sum_m \varepsilon_{mm} = 0$ hold. However, we will find it convenient
to also include the antisymmetric tensor and dilaton components,
thus enlarging the polarization tensor into an arbitrary two index
tensor of SO(4). Contracting this tensor against four dimensional $\sigma$
matrices, the polarization is $\varepsilon_{a \ad; a' \ad'} = \varepsilon_{mn} \sigma^m_{a
\ad} \sigma^n_{a' \ad'}$. Note that in this case the helicity of each
graviton scattering is labeled by four indices $(a, \ad, a', \ad')$.

At this point we invoke the KLT relations \cite{Kawai:1985xq},
which relate amplitudes in gravity to the square of amplitudes in
Yang-Mills theory:
\begin{equation}
\mathcal{M}_3 = A_3 A_3
\end{equation}
where $\mathcal{M}_3$ is the gravitational three point function.
We immediately deduce the simple formula
\begin{equation}
\mathcal{M}_3 (1_{a a' \ad \ad'}, 2_{b b' \bd \bd'}, 3_{c c' \cd
\cd'}) =- \Gamma_{abc} \Gamma_{a'b'c'} \t \Gamma_{\ad \bd \cd}
\tilde\Gamma_{\ad' \bd' \cd'}.
\end{equation}
This equation describes scattering of all possible polarizations of
gravitons in six dimensions; reducing to four dimensions, we can deduce
expressions for gravitons interacting with gauge bosons and so on.

\section{The BCFW Recursion Relations}

With the three point amplitudes in hand, it is straightforward to
construct all higher point amplitudes via the BCFW recursion relations. In
this section we briefly review these relations and describe an efficient
computational method appropriate in dimensions greater than four.
We then express the recursion relations in the language of the six
dimensional spinor-helicity formalism.

\subsection{A Review of BCFW}

The BCFW recursion relations are an expression for on-shell
amplitudes in terms of sums of products of lower point on-shell
amplitudes evaluated at complex momenta. First proposed for
tree-level YM amplitudes \cite{Britto:2004ap,Britto:2005fq},
recursion relations were later derived for gravity
\cite{Cachazo:2005ca,Bedford:2005yy} and eventually found to be a
quite generic property of tree amplitudes in quantum field
theories in arbitrary dimensions
\cite{ArkaniHamed:2008yf,Cheung:2008dn}. The basic idea of the
recursion relations is to analytically continue two of the
external momenta, $p_1$ and $p_2$, of an amplitude by a complex
parameter, $z$:
\be \hat p_1 &=& p_1 + z q \\
\hat p_2 &=& p_2 - z q \ee
where $q^2=p_1 \cdot  q = p_2 \cdot q = 0$. Note that $\hat{p}_{1,2}$ are
complex but still on-shell, and $q$ has the properties of a polarization
vector. Since the amplitude $A$ is a rational function of the momenta,
it is also an analytic function of $z$. Now as long as $A(z)$ vanishes
appropriately at large $z$, it is entirely defined by the residues at its
poles. As argued in \cite{ArkaniHamed:2008yf}, this asymptotic behavior of
$A(z)$ is true in Yang-Mills theory (gravity) as long as the polarization
of particle 1 is $\varepsilon_1^\mu = q^\mu$ ($\varepsilon^{\mu\nu}_1
= q^\mu q^\nu$), while the polarization of particle 2 is arbitrary.
In this case the residue at each pole in $z$ corresponds to a product of
two on-shell lower point amplitudes evaluated at complex momenta, yielding
the following formula for amplitudes in Yang-Mills theory and gravity:
\be A(p_1,p_2,\ldots) &=& \sum_{L,R} \sum_{h,h'} \left( \frac{i P_{h,h'}}{k^2} \right) A_L(\hat
p_1(z_*),\ldots, \hat k(z_*); h) A_R(\hat p_2(z_*),\ldots,
-\hat k(z_*); h') \bigg|_{z_*= - 2 q \cdot p_L} \ee
where $\hat k(z_*)$ and $k$ denoted the the shifted momentum and
physical momentum of the intermediate leg, respectively, while $h$ and
$h'$ labels its polarizations, the ellipsis $\cdots$ denotes the other
external momenta, and $L,R$ sums over partitions of the external legs
into two groups. The operator $P_{h,h'}$ is the sum over a complete set
of propagating states which occurs in the numerator of a propagator.

\subsection{Covariantizing the Recursion Relation}

Ultimately, our goal is to compute a matrix of amplitudes whose
matrix elements correspond to each possible choice of the external
polarizations. In conventional field theory the nearest
approximation to this is the usual amplitude, $A_{\mu_1 \mu_2}$,
where $\mu_1$ and $\mu_2$ are dotted into the polarization
six-vectors for particles 1 and 2. However, as we know, this
particular representation is gauge redundant. Instead, we want
the object $A_{h_1 h_2} = v_{h_1}^{\mu_1}
 A_{\mu_1 \mu_2}
v_{h_2}^{\mu_2}$, where $v$ is a basis for the external
polarization states labeled by a little group index $h_{1,2}$. For
example, in the case of four dimensions this index labels
helicity, so $h_{1,2} = \pm$. Consequently, the matrix elements
of $A_{h_1 h_2}$ correspond to every combination of helicities for
particles 1 and 2: $A_{--}$, $A_{-+}$, $A_{+-}$ and $A_{++}$.
Likewise, in six dimensions $h_1$ labels $(a,\ad)$ indices, $h_2$
labels $(b,\bd)$ indices, etc.

Unfortunately, conventional BCFW is poorly equipped to evaluate $A_{h_1
h_2}$ because it only applies when the deformation vector, $q$, is
chosen to be equal to the polarization of particle 1, $q=\varepsilon_1$.
However, this $q$ enters ubiquitously into the right hand side of the BCFW
reduction---thus to evaluate $A_{h_1 h_2}$ it would be necessary to apply
the recursion relations for every linearly independent choice of $q$!
Luckily, there is a simple way around this, which is to choose $q=X_h
v_h$ to be an arbitrary linear combination of the $v$'s labeled by a
little group vector $X$; we then use the recursion relations to compute
$X_h A_{h,h_2, \ldots}$, the amplitude with appropriate polarization of
particle 1. This is the same as the usual BCFW shift except that we are
keeping the deformation direction unspecified---as such, the recursion
relations do not manifestly break little group covariance. A key point
is that the result of the computation $ X_h A_{h,h_2, \ldots}$ is linear
in $X_h$; after all, this result is simply the amplitude for particles
scattering with particle 1, where particle 1 is in the polarization
state $X_h v_h$. Therefore, it is straightforward to deduce the full
amplitude $A_{h_1,h_2, \ldots}$ as the coefficient of $X_h$. We will
demonstrate this procedure in examples below. As a final comment, we
note that in order to keep $\hat p_{1,2}$ on-shell, we must demand that
$q^2 = X_h X_{h'} (v_h  \cdot v_{h'}) = 0$.

\subsection{Application to Six Dimensions}

Thus far, the discussion of the BCFW recursion relations have been
independent of spacetime dimensionality; in this section, we
specialize to six dimensions and introduce some notation that we
will use to compute the four and five point amplitudes below. We
begin with a simplifying choice of gauge: we take the gauge of
particle 1 to be $p_2$. Then our modified BCFW deformations
become
\begin{align}
\label{eq:vecshifts}
\hat{p}_1 &= p_1 + z X^{a \ad} \pol 1 a \ad\\
\hat{p}_2 &= p_2 - z X^{a \ad} \pol 1 a \ad.
\end{align}
where the on-shell constraint, $\hat{p}_{1,2}^2=0$, fixes $X^{a \ad}X^{b
\bd} \epsilon_{ab}\epsilon_{\ad \bd} = 2 \det X = 0$. Since $X$ has zero
determinant, it is convenient to express it as $X^{a \ad} = x^a \t x^\ad$
and to define
\begin{equation}
y^b = \tilde x^\ad \langle 2_b | 1^\ad]^{-1}, \quad \tilde y_\bd =
x^a \langle 1^a | 2_\bd ]^{-1}.
\end{equation}
Then we find that we can implement the vectorial shifts in
Eq.~\eqref{eq:vecshifts} by the spinorial shifts
\begin{align}
|\hat 1_a \rangle &= |1_a \rangle + z x_a | y \rangle \\
|\hat 2_b \rangle &= |2_b \rangle + z y_b | x \rangle \\
|\hat 1_\ad ] &= |1_\ad ] - z \t x_\ad | \t y ] \\
|\hat 2_\bd ] &= |2_\bd ] - z \t y_\bd | \t x ] ,
\end{align}
where $|x \r = x^a | 1_a \r$ and so on.
The BCFW recursion relations for gauge theory can then be written
\begin{equation}
x^a \t x^\ad A_{a \ad b \bd \ldots}(p_1, p_2, \ldots) = \sum_{L,R}
\sum_{c \cd} \left( -\frac{i}{k^2} \right) x^a \t x^\ad A_{a \ad c \cd}(\hat p_1(z^*), \ldots,
\hat k) A_{b \bd}^{\;\;\;\;c \cd} (\hat p_2(z^*), \ldots, - \hat k),
\end{equation}
where $k$ and $(c,\cd)$ are the momentum and polarization of the
intermediate leg. It is worth noticing that a single BCFW computation in
six dimensions allows one to deduce results for the scattering of particles
in all possible helicity states in four dimensions by a simple dimensional
reduction.

\section{Computing the Four Point Amplitude}

\begin{figure}[t]
\centering
\includegraphics{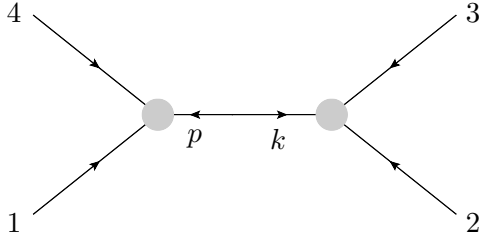}
\put(-60,26){$4$}
\put(1,26){$3$}
\put(1,-2){$2$}
\put(-60,-2){$1$}
\put(-36, 10){$p$}
\put(-25, 9){$k$}
\caption{BCFW diagram for the four point amplitude.}\label{fig:diagram1}
\end{figure}

It is now rather easy to use BCFW recursion to compute a compact formula
for the four point amplitudes of gauge theory and gravity. We choose to
shift the momenta of particles 1 and 2; then there is one BCFW diagram,
Figure~\ref{fig:diagram1}.
The four point function is given by
\begin{equation}
x^a \t x^\ad A_{4; a \ad b \bd c \cd d \dd} = \frac{i}{t} x^a \t x^\ad A_{L; a \ad e \ed d \dd} A_{R; b \bd c \cd}{}^{e \ed}
\end{equation}
where $A_L$, $A_R$ are the left- and right-hand three point function in
the figure, respectively. Since the three point amplitudes are products
of dotted and undotted tensors we can focus our discussion on the undotted
tensor. We must compute
\begin{equation}
\Gamma_{L; a e d} \Gamma_{R; b c}{}^e = ( u_{1a} u_{p e} w_{4 d} + u_{1a} w_{p e} u_{4 d} + w_{1a} u_{p e} u_{4 d} ) ( u_{2b} u_{3 c} w_k^e + u_{2b} w_{3 c} u_k^e + w_{2b} u_{3c} u_k^e).
\end{equation}
It is helpful to choose the spinors associated with the momentum $p =
-k$ to be $|p] = i |k]$ and $|p\rangle = i |k \rangle$. Then we find
three key properties of the $u$ and $w$ spinors associated with $p$
and $k$. These are, firstly, that
\begin{equation}
\tilde u_p \cdot \tilde u_k u_p \cdot u_k = \tilde u_p^{\ed} \tilde u_{k \ed} u_p^e u_{p e} = - s,
\end{equation}
so that, in particular, $u_p \cdot u_k \neq 0$. Consequently, we can
exploit the $b$ redundancy of $w_k$ and $w_p$ to choose, secondly,
\begin{equation}
u_p \cdot w_k = \t u_p \cdot \t w_k = w_p \cdot u_k = \t w _p \cdot \t u_k = 0.
\end{equation}
For, if $u_p \cdot w_k \neq 0$, for example, then we can choose $w'_k
= w_k + b_k u_k$ such that $u_p \cdot w'_k = u_p \cdot w_k + b_k u_p
\cdot u_k = 0$. This equation always has a solution for $b_k$ since
$u_p \cdot u_k \neq 0$. Finally, it is easy to show that
\begin{equation}
w_k \cdot w_p = \frac{1}{u_k \cdot u_p}.
\end{equation}
In view of these three properties, we conclude that we can choose
normalizations so that
\begin{equation}
\tilde w_{k} = \frac{ \t u_p}{\sqrt{-s}}, \quad w_k = \frac{u_p}{\sqrt{-s}}.
\end{equation}
The undotted tensorial part of the amplitude now simplifies to
\begin{multline}
\Gamma_{L; a e d} \Gamma_{R; b c}{}^e = \frac{1}{u_k \cdot u_p} \left[ u_{1a} u_{2b} u_{3c} u_{4d} - s \left( u_{1a} u_{2b} w_{3c} w_{4d} \right . \right. \\
\left. \left. +  u_{1a} w_{2b} u_{3c} w_{4d}  +  w_{1a} u_{2b} w_{3c} u_{4d} +  w_{1a} w_{2b} u_{3c} u_{4d} \right) \right].
\end{multline}
However, it is easy to see that
\begin{equation}
\langle \hat 1_a \hat 2_b 3_c 4_d \rangle =  \left[ u_{1a} u_{2b} u_{3c} u_{4d} - s \left( u_{1a} u_{2b} w_{3c} w_{4d}
+  u_{1a} w_{2b} u_{3c} w_{4d}  +  w_{1a} u_{2b} w_{3c} u_{4d} +  w_{1a} w_{2b} u_{3c} u_{4d} \right) \right].
\end{equation}
For example, we compute
\begin{align}
u_1^a u_2^b \langle \hat 1_a \hat 2_b 3_c 4_d \rangle &= i \langle u_p \cdot k  \; u_k \cdot k \; 3_c 4_d \rangle \\
&= -i u_p \cdot u_k \langle 3_c | \ks | 4_d \rangle \\
&= - u_p \cdot u_k \; u_{3c} \t u_k^e \; [p_e | 4_d \rangle \\
&= - s u_{3c} u_{4d}.
\end{align}
All other components of $ \langle \hat 1_a \hat 2_b 3_c 4_d \rangle$
can be projected onto the $u_i$, $w_i$ basis of the tensor product space
in the same way. Thus, we find
\begin{equation}
x^a \t x^\ad A_{4; a \ad b \bd c \cd d \dd} = \frac{-i}{st} x^a \t x^\ad \langle \hat 1_a \hat 2_b 3_c 4_d \rangle [\hat 1_\ad \hat 2_\bd 3_\cd 4_\dd ] = \frac{-i}{st} x^a \t x^\ad \langle 1_a 2_b 3_c 4_d \rangle [ 1_\ad 2_\bd 3_\cd 4_\dd ].
\end{equation}
Since this final expression is manifestly linear in $x$ and $\t x$,
we can deduce that
\begin{equation}
 A_{4; a \ad b \bd c \cd d \dd} = \frac{-i}{st}  \langle 1_a 2_b 3_c 4_d \rangle [ 1_\ad 2_\bd 3_\cd 4_\dd ].
\end{equation}

With the expression for the Yang-Mills four point function in hand, it
is a trivial matter to deduce the gravitational four point function. The
KLT relation in this case is
\begin{equation}
\mathcal M_4(1,2,3,4) = -i s A_4(1,2,3,4) A_4(1,2,4,3),
\end{equation}
so we immediately deduce that
\begin{equation}
\label{eq:grav4pt}
\mathcal M_4(1,2,3,4) = \frac{i}{s t u} \langle 1_a 2_b 3_c 4_d \rangle \langle 1_{a'} 2_{b'} 3_{c'} 4_{d'}  \rangle [  1_\ad 2_\bd 3_\cd 4_\dd ] [ 1_{\ad'} 2_{\bd'} 3_{\cd'} 4_{\dd'} ].
\end{equation}
The compactness of this explicit expression for the gravitational four
point amplitude is an illustration of the power of the spinor-helicity
formalism. Of course, this occurs simply because these variable capture
physical properties of the single particle state with no redundancy.

\section{Five Points}\label{sec:fivepoint}

The final amplitude we will discuss in this work is the five point
amplitudes for Yang-Mills theory. We will compute the five point function using the BCFW recursion
relations; then the KLT relations can by used to deduce the gravitational amplitude.

\begin{figure}[h]
\centering
\subfigure[]{
\includegraphics[scale=1.4]{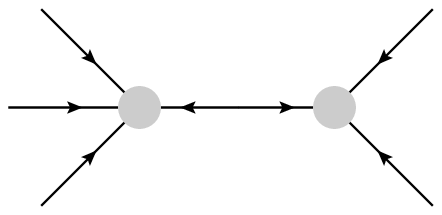}
\put(-64,13.5){$5$}
\put(-60,28){$4$}
\put(1,28){$3$}
\put(1,-2){$2$}
\put(-60,-2){$1$}
\put(-36, 11){$p$}
\put(-23, 10){$k$}
}
\hspace{0.4in}
\subfigure[]{
\includegraphics[scale=1.4]{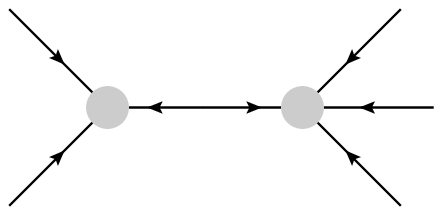}
\put(-64,28){$5$}
\put(-4,28){$4$}
\put(0, 13){$3$}
\put(-4,-2){$2$}
\put(-64,-2){$1$}
\put(-41,11){$p$}
\put(-28,10){$k$}
}
\caption{BCFW diagrams for the five point amplitude.}\label{fig:5pt}
\end{figure}

As in our discussion of the four point amplitude, we choose to shift the momenta of particles 1 and 2. There are now two BCFW diagrams, shown in Figure~\ref{fig:5pt}. Using the identity
\begin{equation}
|k_e\r \Gamma_{bc}{}^e= -u_c |\hat 2_b \r - u_b |3_c \r,
\end{equation}
the first diagram can be written as
\begin{equation}
D_1 = \frac{-i}{s_{45} \hat s_{51} s_{23}} \left( \l \hat 1_a \hat 2_b 4_d 5_e \r u_c + \l \hat 1_a \hat 3_c 4_d 5_e \r u_b  \right) \left(  [ \hat 1_\ad \hat 2_\bd 4_\dd 5_\ed ] \t u_\cd +[ \hat 1_\ad \hat 3_\cd 4_\dd 5_\ed ] \t u_\bd \right),
\end{equation}
where $s_{ij} = (p_i + p_j)^2$ and $\hat s_{51} = (\hat p_1 + p_5)^2$. Since the first step in these calculations is to express the sum of diagrams (appropriately contracted with $x \tilde x$) in a form which is manifestly linear in $x$ and $\t x$, our first goal in simplifying each diagram is to make the $x$ and $\t x$ dependence as clear and as simple as possible. In this vein, we define $z q \equiv \hat p_1(z) - p_1$, and study $\hat s_{15}$. Notice that
\begin{equation}
\hat s_{15} q \cdot p_3 = s_{15} q \cdot p_3 + s_{23} q \cdot p_5
= \frac{1}{2 s_{12}} \l x | \ps_5 \ps_4 \ps_3 \ps_2 | \t x] \equiv x^a \t x^\ad \phi_{a \ad}.
\end{equation}
Furthermore, we find that
\begin{equation}
q \cdot p_3 = \frac{1}{2 s_{12}} \l x | 2^\bd] \t u_{2 \bd} u_{2 b} [\t x| 2^\b \r.
\end{equation}
Putting these results together, and contracting in with $x^a \t x^\ad$, we find
\begin{multline}
X \cdot D_1 = \frac{i}{2  (x \cdot \phi\cdot  \t x) s_{12} s_{23} s_{45}} \left( - \l x 2_b 4_d 5_e \r \l x | \ps_2 | 3_c \r + \l x 3_c 4_d 5_e \r \l x | \ps_3 | 2_b \r  \right) \\
\times \left( - [ \t x 2_\bd 4_\dd 5_\ed ] [ \t x | \ps_2 | 3_\cd ] + [ \t x 3_\cd 4_\dd 5_\ed ] [ \t x | \ps_3 | 2_\bd ]  \right).
\end{multline}
Similarly, we find for the second diagram in the figure,
\begin{multline}
X \cdot D_2 =  \frac{i}{2 (x \cdot \phi\cdot  \t x) s_{12}^2 s_{34} s_{15}} \left( \l 2_b 3_c 4_d 5_e \r \l x | \ps_5 \ps_2 | \t x] - s_{12} \l 2_b 3_c 4_d x \r \l 5_e | \t x] + s_{15} \l2_b | \t x] \l x 3_c 4_d 5_e \r \right) \\
\times \left( [ 2_\bd 3_\cd 4_\dd 5_\ed ] [ \t x | \ps_5 \ps_2 | x\r - s_{12} [ 2_\bd 3_\cd 4_\dd \t x ] [ 5_\ed | \t x\r + s_{15} [2_\bd | \t x\r [ \t x 3_\cd 4_\dd 5_\ed ] \right).
\end{multline}
Notice that neither of the diagrams is linear in $x$ or $\t x$. However, their sum is linear as expected. This is most easily seen by using the Schouten identity, Eq.~\eqref{eq:schouten} to rewrite the first diagram in the form
\begin{multline}
X \cdot D_1 = \frac{i}{2  (x \cdot \phi\cdot  \t x) s_{12} s_{23} s_{45}} \left( \l x 2_b 3_c 4_d \r \l x | \ps_4 | 5_e \r - \l x 2_b 3_c 5_e \r \l x | \ps_5 | 4_d \r \right) \\
\times  \left( [ \t x 2_\bd 3_\cd 4_\dd ] [\t x | \ps_4 | 5_\ed ] - [ \t x 2_\bd 3_\cd 5_\ed ] [ \t x | \ps_5 | 4_\dd ] \right).
\end{multline}
Meanwhile, another use of the Schouten identity allows us to remove some of the $x$ dependence in the denominator of the second diagram. In particular, we can write the dotted tensor structure, for example, of diagram 2 as
\begin{multline}
\l 2_b 3_c 4_d 5_e \r \l x | \ps_5 \ps_2 | \t x] - s_{12} \l 2_b 3_c 4_d x \r \l 5_e | \t x] + s_{15} \l2_b | \t x] \l x 3_c 4_d 5_e \r = - \frac{2 (x \cdot \phi \cdot \t x)}{s_{23}} \l 2_b 3_c 4_d 5_e \r \\
+ \l 2_b 3_c 4_d x \r \frac{[\t x | \ps_2 \ps_3 \ps_4 \ps_1 | 5_e\r}{s_{12} s_{23}} + \l 5_e x 2_b 3_c \r \frac{[\t x| \ps_5 \ps_1 \ps_2 \ps_3|4_d \r}{s_{12}s_{23}}.
\end{multline}
To complete the cancellation of the quantity $(x \cdot \phi \cdot \t x)$, we simply use the rearrangement formulae given in Appendix~\ref{sec:rearrange}.
Some further use of the Schouten identity then yields an expression for the five point function which is most conveniently described in terms of two tensors:
\begin{equation}
\mathcal{M}_{a \ad b \bd c \cd d \dd e \ed} = \frac{1}{s_{12} s_{23} s_{34} s_{45} s_{51}} \left( \mathcal{A}_{a \ad b \bd c \cd d \dd e \ed} + \mathcal{D}_{a \ad b \bd c \cd d \dd e \ed}\right) ,
\end{equation}
where
\begin{equation}
\mathcal{A}_{a \ad b \bd c \cd d \dd e \ed} = \l 1_\a | \ps_2 \ps_3 \ps_4 \ps_5| 1_\ad ] \l 2_b 3_c 4_d 5_e \r \lbrack 2_\bd 3_\cd 4_\dd 5_\ed \rbrack + \textrm{ cyclic permutations,}
\end{equation}
and
\begin{multline}
\mathcal{D}_{a \ad b \bd c \cd d \dd e \ed} = \l 1_a {(2 . \Delta_2)_\bd}] \l 2_b 3_c 4_d 5_e \r [1_\ad 3_\cd 4_\dd 5_\ed ]
+\l 3_c {(4 . \Delta_4)_\dd}] \l 1_a 2_b 4_d 5_e \r [1_\ad 2_\bd 3_\cd 5_\ed ] \\
+ \l 4_d {(5 . \Delta_5)_\ed}] \l 1_a 2_b 3_c 5_e \r [1_\ad 2_\bd 3_\cd 4_\dd ]
+ \l 3_c {(5 . \Delta_5)_\ed}] \l 1_a 2_b 4_d 5_e \r [1_\ad 2_\bd 3_\cd 4_\dd ],
\end{multline}
where the matrices $\Delta_i$ are defined by
\begin{equation}
\Delta_1 = \l 1 | \ps_2 \ps_3 \ps_4 - \ps_4 \ps_3 \ps_2 |1 ],
\end{equation}
with the other $\Delta_i$ defined by cyclic permutations of this formula. Notice that, while the tensor $\mathcal{A}_{a \ad b \bd c \cd d \dd e \ed}$ is
manifestly symmetric under cyclic permutations of the particle label, $\mathcal{D}_{a \ad b \bd c \cd d \dd e \ed}$ does not obviously have this
symmetry. However, it is easy to see that it is symmetric using the Schouten identity.

The gravitational amplitude can then be obtained using the KLT relation. In the case of a five point amplitude, this relation is
\begin{equation}
\mathcal M_5 = s_{23} s_{45} A(1,2,3,4,5) A(1,3,2,5,4) + s_{24} s_{35} A(1,2,4,3,5) A(1,4,2,5,3).
\end{equation}
It is now a matter of algebra to deduce an expression for the gravitational five point amplitude.

\section{Concluding Remarks}

The main achievement of this work has been to find a viable
spinor-helicity formalism in six dimensions. That this formalism
has the potential to be useful is clear from the simplicity of
amplitudes in this framework. In particular, it is remarkable that
we now have an explicit, gauge independent, compact formula for
the gravitational four point amplitude, given in
Eq.~\eqref{eq:grav4pt}. Since it has been possible to extend the
spinor-helicity formalism from four to six dimensions, it is fair
to ask whether the same is possible for even higher dimensions. A
formalism in ten dimensions, for example, might be particularly
interesting from the point of view of Yang-Mills theory.

We believe that scattering amplitudes take a remarkable simple
form in terms of spinors because these variables encode precisely
the physical degrees of freedom of asymptotic states. In
particular, amplitudes expressed as functions of spinors transform
appropriately under the little group without the need for an
unphysical gauge redundancy. That is, the success of the formalism
has a physical motivation---it is not a mathematical trick.

In the arena of six dimensions, there are many interesting questions
that presently are unanswered. Parke and Taylor~\cite{Parke:1986gb}
wrote down a compact formula for $n$-point MHV scattering amplitudes
in four dimensions. This class of $n$-point amplitudes is particularly
simple, so one might want to examine a simplified subset of amplitudes in
six dimensions. However, it is impossible to find such a subset which is closed
under Lorentz transformations, because all of the
polarization states in six dimensions are connected by a continuous
SO(4) symmetry and there is no conserved helicity. The flip side of this statement is that any expression
for the $n$ point amplitudes in six dimensions would amount to complete
knowledge of the tree-level $S$-matrix - such an expression would be an
exciting discovery.

Finally, while we have presented results for gravity and gauge
theory, there are other theories in six dimensions which we have
left untouched. One particularly interesting theory is the $(2,0)$
theory~\cite{Witten:1995zh}, about which rather little is known. Perhaps insight into
this theory might be obtained using these novel kinematic
variables. It would also, of course, be interesting to investigate
supersymmetric theories.

\appendix

\section{The Clifford Algebra}\label{sec:clifford}

Let us start at the beginning. We work with the mostly negative
metric, and define Pauli matrices
\begin{equation}
\sigma_0 = \begin{pmatrix}
1 & 0 \\
0 & 1
\end{pmatrix},
\;\; \sigma_1 = \begin{pmatrix}
0 & 1 \\
1 & 0
\end{pmatrix},
\;\; \sigma_2 = \begin{pmatrix}
0 & -i \\
i & 0
\end{pmatrix},
\;\; \sigma_3 = \begin{pmatrix}
1 & 0 \\
0 & -1
\end{pmatrix}.
\;\;
\end{equation}
The Clifford algebra is
\begin{equation}
\sigma^\mu \tilde \sigma^\nu + \sigma^\nu \tilde \sigma^\mu = 2
\eta^{\mu \nu}.
\end{equation}
We will work with a particular basis of this algebra. The Lorentz
group SO(6) is isomorphic to SU(4); the spinors of SO(6) are the
fundamentals of SU(4). The antisymmetric tensor of SU(4) is the
fundamental of SO(6). Therefore, we can choose a basis of the
Clifford algebra so that $\sigma, \;\; \tilde \sigma$ are
antisymmetric. At the same time, it is convenient to work with a
basis which is simply related to a standard choice of $\gamma$
matrices in four dimensions. Our choice is
\begin{subequations}
\begin{align}
\sigma^0 &= i \sigma_1 \otimes \sigma_2 & \tilde \sigma^0 &= -i \sigma_1 \otimes \sigma_2 \\
\sigma^1 &= i \sigma_2 \otimes \sigma_3 & \tilde \sigma^1 &= i
\sigma_2 \otimes
\sigma_3 \\
\sigma^2 &= - \sigma_2 \otimes \sigma_0 & \tilde \sigma^2 &=
\sigma_2 \otimes
\sigma_0 \\
\sigma^3 &= -i \sigma_2 \otimes \sigma_1 & \tilde \sigma^3 &= -i \sigma_2 \otimes \sigma_1 \\
\sigma^4 &= - \sigma_3 \otimes \sigma_2 & \tilde \sigma^4 &= \sigma_3 \otimes \sigma_2 \\
\sigma^5 &= i \sigma_0 \otimes \sigma_2 & \tilde \sigma^5 &= i
\sigma_0 \otimes \sigma_2.
\end{align}
\end{subequations}
We adopt the convention that the six
dimensional $\sigma^\mu$ have lower indices while the $\tilde
\sigma^\mu$ have upper indices. These objects enjoy the properties
\begin{align}
\label{eq:sigmaprops}
\sigma^\mu_{AB} \sigma_{\mu CD} &= -2 \epsilon_{ABCD},\\
\tilde \sigma^{\mu AB} \tilde \sigma_{\mu}^{CD} &= -2 \epsilon^{ABCD}, \\
\sigma^\mu_{AB} \tilde \sigma_{\mu}^{CD} &= -2 \left( \delta_A^C
\delta_B^D
-\delta_A^D \delta_B^C \right), \\
\tr \sigma^\mu \tilde \sigma^\nu &= 4 \eta^{\mu \nu},
\end{align}
where $\epsilon_{1234} = \epsilon^{1234} = 1$. 

The final identity we will discuss in this appendix is the six-dimensional generalization of the Schouten identity.
Since the spinors live in a six dimensional space, linear dependence of five (chiral) spinors implies
\begin{equation}
\label{eq:schouten}
\l 1234 \r \l 5| + \l 2345 \r \l 1| + \l 3451 \r \l 2| + \l 4512 \r \l 3| + \l 5123 \r \l 4| =0.
\end{equation}
Of course, a similar equation holds for anti-chiral spinors.

\section{Rearrangement Formulae}\label{sec:rearrange}

These rearrangement formulae are useful for simplifying the sum of the two diagrams encountered in the computation of the 5 point amplitude described in section~\ref{sec:fivepoint}. In the notation we used in our discussion of the five point amplitude, the following identities hold:
\begin{align}
&\frac{1}{s_{12}^2 s_{23}^2 s_{34} s_{15}} [\t x| \ps_2 \ps_3 \ps_4 \ps_1 | 5_e \r \l x | \ps_2 \ps_3 \ps_4 \ps_1 | 5_\ed ] + \frac{1}{s_{12}s_{23}s_{45}} \l x |\ps_4 |5_e \r [\t x| \ps_4| 5_\ed] \nonumber\\
&\quad= 2 \frac{x \cdot \phi \cdot \t x} {s_{12}s_{23}^2 s_{34} s_{45} s_{51}} \l 5_e| \ps_1 \ps_4 \ps_3 \ps_2 \ps_1 \ps_4 | 5_\ed],\\
%
&\frac{1}{s_{12}^2 s_{23}^2 s_{34} s_{15}} [\t x| \ps_5 \ps_1 \ps_2 \ps_3 | 4_d \r \l x | \ps_5 \ps_1 \ps_2 \ps_3 | 4_\dd ] + \frac{1}{s_{12}s_{23}s_{45}} \l x |\ps_5 |4_d \r [\t x| \ps_5| 4_\dd] \nonumber\\
&\quad= 2 \frac{x \cdot \phi \cdot \t x} {s_{12}s_{23}^2 s_{34} s_{45} } \l 4_d |\ps_5 \ps_1 \ps_2 \ps_3 | 4_\dd],\\
%
&\frac{1}{s_{12}^2 s_{23}^2 s_{34} s_{15}} [\t x| \ps_5 \ps_1 \ps_2 \ps_3 | 4_d \r [\t x | \ps_2 \ps_3 \ps_4 \ps_1 | 5_e \r - \frac{1}{s_{12}s_{23}s_{45}} \l x |\ps_5 |4_d \r [\t x| \ps_4| 5_e] \nonumber\\
&\quad= 2 \frac{x \cdot \phi \cdot \t x} {s_{12}s_{23}^2 s_{34} s_{45} } \l 4_d| \ps_5 \ps_1 \ps_2 \ps_3  \ps_4 \ps_1| 5_\ed].
\end{align}

\acknowledgements

DOC is the Martin A. and Helen Chooljian Member at the Institute for
Advanced Study, and was supported in part by the US Department
of Energy under contract DE-FG02-90ER-40542. We thank 
Nima Arkani-Hamed, Henriette Elvang, Gil Paz, Jared Kaplan, Michele 
Papucci and Brian Wecht for useful discussions.

\end{document}